\documentclass[twocolumn,pra,floatfix]{revtex4}

\usepackage{amsmath}
\usepackage{amssymb}
\usepackage{bm}                          
\usepackage{ifthen}
\usepackage{color}
\usepackage{graphicx}

\DeclareMathAlphabet{\matheurm}{U}{eur}{m}{n}

\newcommand\eq[1]                              
{
\begin{align*}
#1
\end{align*}
}

\newcommand\eql[2] 
{
\begin{equation}\label{#1}
\begin{split}
#2
\end{split}
\end{equation}
}

\newcommand\eqnl[2]        
{
\begin{equation}\label{#1}
#2
\end{equation}
}

\newcommand\eqsl[1]                            
{
\begin{align}
#1
\end{align}
}

\newcommand\eqssl[2]                      
{
\begin{subequations}\label{#1}
\begin{align}
#2
\end{align}
\end{subequations}
}

\renewcommand\_[1]     {_{#1}^{}}

\providecommand\eqref[1] {(\ref{#1})}  
\newcommand\Eq[1]      {Eq.~\eqref{#1}}
\newcommand\Eqs[1]     {Eqs.~\eqref{#1}}
\newcommand\Fig[1]     {Fig.~\ref{#1}}
\newcommand\Sec[1]     {Sec.~\ref{#1}}
\newcommand\Ref[1]     {Ref.~\onlinecite{#1}}

\newcommand\Cite[1]    {~\cite{#1}}                

\newcommand\ME[3]      {\langle{{#1}}|{{#2}}|{{#3}}\rangle} 
\newcommand\ket[1]     {|{{#1}}\rangle}

\newcommand\braket[2]  {\langle{{#1}}|{{#2}}\rangle}

\newcommand\Matrix[1]                    
{{\begin{pmatrix} #1 \end{pmatrix}}}

\newcommand\matrixgHC[2][]                           
                       {{\bm{#2}#1^\dagger}}

\newcommand\Db[1]      {b_{{#1}}^{}}
\newcommand\Cb[1]      {b_{{#1}}^\dagger}
\newcommand\Dc[1]      {c_{{#1}}^{}}
\newcommand\Cc[1]      {c_{{#1}}^\dagger}

\newcommand\Dvarphi[1] {\hat{\varphi}_{{#1}}^{}}
\newcommand\Cvarphi[1] {\hat{\varphi}_{{#1}}^\dagger}
\newcommand\Dpsi[1]    {\hat{\psi}_{{#1}}^{}}
\newcommand\Cpsi[1]    {\hat{\psi}_{{#1}}^\dagger}

\newcommand\Pgs        {\mathcal{P}_\mathrm{gs}}      

\newcommand\Aop        {{\hat{A}}}
\newcommand\Bop        {{\hat{B}}}

\newcommand\HSop       {{\hat{v}}}
\newcommand\Hop        {{\hat{H}}}
\newcommand\HGP        {{\hat{H}_{\mathrm{GP}}}}
\newcommand\Kop        {{\hat{K}}}
\newcommand\Top        {{\hat{T}}}
\newcommand\Vop        {{\hat{V}}}

\newcommand\XV[1]      {\langle{#1}\rangle}

\newcommand\kinXV      {\XV{\Top}}
\newcommand\trapXV     {\XV{\Vop_{\mathrm{trap}}}}
\newcommand\potXV      {\XV{\Vop_{\mathrm{2B}}}}

\newcommand\BPEst[1]   {{#1}_{\mathrm{bp}}}

\newcommand\half       {{\textstyle\frac{1}{2}}}
\newcommand\Half       {\frac{1}{2}}

\newcommand\intdr[1][] {\! \int \! d^3\rvec#1 \,}

\newcommand\expP[1]    {\exp \! \left( {{#1}} \right)} 

\newcommand\aho[1][]   {a_{\mathrm{ho}}^{#1}}            
\newcommand\as[1][]    {a_s^{#1}}              
\newcommand\ET         {E_\mathrm{T}}                       
\newcommand\EL         {E_\mathrm{L}^{}}         
\newcommand\EBog       {E_\mathrm{Bog}}      
\newcommand\PsiT[1][]  {\Psi_{\mathrm{T}#1}^{}}

\newcommand\PsiGS      {\Phi_0}

\newcommand\Dt         {\Delta\tau}                  
\newcommand\tauBP      {\tau_{\mathrm{bp}}^{}}      
\newcommand\expdtK     {\exp \! {\big( \! - \! \half \Dt \Kop \big)}}

\newcommand\expdtV     {\exp \! {\big( \! - \! \Dt \Vop \big)}}

\newcommand\exptBPH    {\exp \! {\big( \! - \! \tau_{\mathrm{bp}}^{}
                                       \Hop \big)}}
\newcommand\Order      {\mathcal{O}}
\newcommand\spin[1]    {#1}       
\newcommand\spinsum[1] {\sum_{#1}}                             

\newcommand\sigmabar[1][]  {\bar{\sigma}_{#1}}

\newcommand\xsqrbar[1][] 
                       {\underline{{x}_{#1}^2}}

\providecommand\I{}                         
\renewcommand\I{\mathrm{i}}               

\newcommand\wlkridx[1] {\matheurm{#1}}               

\newcommand\wIDX[3][]  {
                        \ifthenelse{\equal{#3}{}}
                          {\ifthenelse{\equal{#1}{}}
                            {_{\wlkridx{#2}}^{}}
                            {_{\wlkridx{#2}}#1}
                          }
                          {_{\wlkridx{#2}}^{(#3)}{}#1}
                       }
\newcommand\wlkr[3][]  {\phi\wIDX[#1]{#2}{#3}}

\newcommand\Wlkr[3][]  {\ket{\wlkr[#1]{#2}{#3}}}

\newcommand\Ovlp[3][]  {w\wIDX[#1]{#2}{#3}}

\newcommand\bwlkr[3][] {
                        \eta_{\wlkridx{#2}}
                        ^{\ifthenelse{\equal{#3}{}}{}{(#3)}}
                        {}#1
                       }

\newcommand\bOvlp[3][] {
                        u_{\wlkridx{#2}}
                        ^{\ifthenelse{\equal{#3}{}}{}{(#3)}}
                        {}#1
                       }

\newcommand\kvec       {\mathbf{k}}                           
\newcommand\rvec       {\mathbf{r}}                           

\newcommand\Angstrom   {\ensuremath{\,\text{\r{A}}}}
\newcommand\um         {\ensuremath{\mu\mathrm{m}}}

\newcommand\fmtSC[1]   {}
\newcommand\fmtTC[1]   {#1}

\fmtTC{
\renewcommand\expP[1]    {e^{
    \renewcommand\half {\frac{1}{2}}
    #1
}} 
\renewcommand\expdtK     {e^{-\Half \Dt \Kop}}

\renewcommand\expdtV     {e^{-\Dt \Vop}}

\renewcommand\exptBPH    {e^{-\tau_{\mathrm{bp}}^{} \Hop}}
} 

\renewcommand\spin[1]    {}
\renewcommand\spinsum[1] {}

\definecolor{Green}{rgb}{0.2,0.96,0.2}
\definecolor{Remarks}{rgb}{1,0.3,0.3}
\definecolor{Extra}{rgb}{0.2,0.2,1}
\definecolor{Blue}{rgb}{0.2,0.3,1}
\definecolor{Black}{rgb}{0,0,0}

\newcommand\DRAFTWORK[1] {}

\newcommand\COMMENTED[1] {}

\newcommand\PLOTFILE[1]  {plots/#1}     

\begin{document} 

\title{Correlation effects in the ground state
of trapped atomic Bose gases}
\date{\today}

\author{Wirawan Purwanto} \email{wirawan@camelot.physics.wm.edu}
\author{Shiwei Zhang}     \email{shiwei@physics.wm.edu}
\affiliation{Department of Physics,
             The College of William and Mary,
             Williamsburg, Virginia 23187}

\begin{abstract}

We study the effects of many-body correlations
in trapped ultracold atomic Bose gases.
We calculate the ground state of the gas using a ground-state auxiliary-field
quantum
Monte Carlo (QMC) method [Phys.~Rev.~E \textbf{70}, 056702 (2004)].
We examine the properties of the gas, such as the energetics,
condensate fraction, real-space density, and momentum distribution, as a
function of the number of particles and the scattering length.
We find that the mean-field Gross-Pitaevskii (GP) approach gives
qualitatively incorrect result of the kinetic energy as a function
of the scattering length.
We present detailed QMC data for the various quantities, and
discuss the behavior of GP, modified GP, and
the Bogoliubov method under a local density approximation.

\end{abstract}

\COMMENTED{
\begin{verbatim}
TBHQMC PAPER 2: INTERACTION (DRAFT REVISION 9/SUBMISSION 1) ---
CVS $Id: tbh-intx.tex,v 1.16 2005/06/09 04:10:54 wirawan Exp $
---------------------------------------------------------------



\end{verbatim}
}

\maketitle 

\section{Introduction}\label{sec:Introduction} 

The many-body physics in trapped Bose gases has drawn intense interest
since the experimental realization of Bose-Einstein condensation (BEC) in
ultracold, dilute alkali atoms\Cite{Anderson1995}.
The systems are ``clean'' and highly controllable experimentally.
The dominant interactions are simple and well-understood, and
the strength of the interatomic interactions can be readily
tuned by means of Feshbach resonances\Cite{Cornish2000}.
With the recent realization of degenerate Fermi
gases\Cite{DeMarco1999,Schreck2000,Truscott2001},
these ultracold systems provide an ideal
``laboratory'' for studying many-body physics.

In the weakly-interacting regime, mean-field theories work
quite well, for instance the Gross-Pitaevskii (GP)
equation\Cite{Gross1961,Gross1963,Pitaevskii1961} for
boson ground states.
Much work has been done to study the ground state of the Bose atomic gases
beyond mean field.
For example, a modified GP equation was proposed\Cite{Braaten1997}
by inclusion of one-loop quantum corrections
and the use of local-density approximation.
Esry\Cite{Esry1997} developed a Hartree-Fock theory as a means of
including the correlation effects in the BEC many-body calculations.
Mazzanti and co-workers\Cite{Mazzanti2003} applied a correlated basis
theory\Cite{Fantoni1998_BookChapter} to study the detailed structure of
dilute hard- and soft-sphere Bose gases.
A comparative study for the modified GP and correlated basis approaches is
presented in \Ref{Fabrocini1999}.
Recently, McKinney and co-workers\Cite{McKinney2004} used a many-body
dimensional perturbation theory to compute the ground-state energy and
breathing-mode frequency of spherically trapped gases at different
interaction strengths.

Semianalytic methods are versatile and generally very
easy to extend to realistic systems with large number of particles.
However, they are approximate and each has its own
limitations, especially in the strongly-interacting regime.
Computational methods such as
quantum Monte Carlo (QMC) provide a useful, complementary alternative.
A variety of such calculations have been carried out for atomic boson
systems, including
variational Monte Carlo\Cite{DuBois2001}
and
the exact diffusion Monte Carlo (DMC)\Cite{Ceperley1980,Umrigar1993}
studies on
both the homogenous\Cite{Giorgini1999} and
trapped gases\Cite{Blume2001,Sakhel2002,DuBois2003}.

We have recently developed an auxiliary-field
quantum Monte Carlo (AF QMC) method\Cite{Purwanto2004} for the
ground state of many-boson systems.
While the standard DMC works in real space with particle configurations,
our method works in the second-quantized formalism, which
automatically accounts for particle permutation statistics.
The calculation can be carried out in any single-particle basis.
Conceptually, the method provides a way to systematically improve upon
mean field while retaining its basic machinery, capturing correlation
effects with a stochastic, coherent ensemble of independent-particle
solutions.
Various observables and correlation functions can be calculated relatively
straightforwardly.

The initial motivation of this study was to use the AF QMC method to
quantify, by direct comparison with GP, the effects of interactions in
trapped Bose gases, and to provide additional precise numerical data
where they were not available.
(Although the method is not exact for bosons with repulsive
interactions, the systematic errors are very small in the parameter
region of interest, as we show below.)
In particular, we were interested in the behavior of the system as
a function of the interaction strength, which, unlike in typical
condensed matter systems, can be probed directly in experiments.
We found that GP yielded significant errors in the energetics in the
Feshbach resonance regime, which resulted in a qualitatively incorrect
behavior of the kinetic energy in GP as a function of the scattering
length.
To study the origin of these errors, we carried out additional
calculations using first-order Bogoliubov results under a local-density
approximation (LDA).
The purpose of this paper is thus to present our QMC data, and discuss the
behavior of the GP, modified GP, and Bogoliubov-LDA methods as benchmarked
by QMC.

The rest of the paper is organized as follows.
In \Sec{sec:model'}, we describe the many-body Hamiltonian.
Our QMC method is summarized in \Sec{sec:method},
as are the procedures of our GP and Bogoliubov-LDA calculations.
%
Results from QMC, GP, and first-order
Bogoliubov-LDA methods are presented in \Sec{sec:Results},
where we study the energetics of the gas in three dimensions as a function
of the number of particle $N$ and the $s$-wave scattering length $a_s$,
and examine the density profile and momentum distribution.
Our study extends to the strongly-interacting regime achieveable
by Feshbach resonances.
In \Sec{sec:Discussions}, we discuss the implications of our comparisons
between GP, modified GP, Bogoliubov-LDA and QMC. In addition, we
also discuss the influence of the details of the two-body potential.
%
Concluding remarks are given in \Sec{sec:Conclusions}.
Finally, in the appendix,
we describe additional details on our Bogoliubov and QMC
calculations, including benchmark results on the systematic errors
in our QMC.

\section{Modified Bose-Hubbard model}
\label{sec:model'}

We consider $N$ Bose particles in a three-dimensional cube of
side length $2 r\_b$, under the
periodic boundary condition.
Similar to our earlier work\Cite{Purwanto2004}, we use the Bose-Hubbard model
as the discrete representation of the many-body Hamiltonian on a real-space
lattice:
\eql{eq:Hamiltonian_real}
{
    \Hop
& = \frac{\hbar^2}{2m}
    \sum_{\spin{\sigma} \kvec} k^2
    \Cvarphi{\spin{\sigma}}(\kvec) \Dvarphi{\spin{\sigma}}(\kvec)
    \\
& + \;
    \Half m\omega_0^2
    \spinsum{\sigma} 
    \intdr \, r^2 \Cpsi{\spin{\sigma}}(\rvec) \Dpsi{\spin{\sigma}}(\rvec)
    \\
& + \;
    \Half \left( \frac{4\pi a_s \hbar^2}{m} \right)
    \\
& \times \;
    \spinsum{\sigma}
    \intdr[\_1] \intdr[\_2]
    \Cpsi{\spin{\sigma}}(\rvec\_1) \Cpsi{\spin{\sigma}}(\rvec\_2)
    \delta(\rvec\_1 - \rvec\_2)
    \Dpsi{\spin{\sigma}}(\rvec\_2) \Dpsi{\spin{\sigma}}(\rvec\_1)
    \,,
}
where the kinetic energy operator is modified from the Bose-Hubbard form
we used earlier, and is expressed in momentum space instead,
with 
\eql{eq:rq-field-xform}
{
    \Dvarphi{}(\kvec)
& = \frac{1}{(2 r\_b)^{3/2}}
    \int d{\rvec}\, \Dpsi{}(\rvec) e^{\I \kvec \cdot \rvec}
    \,.
}
The sum over $\kvec$ is taken over all the (discretized) momentum
coordinates.
Equation~\eqref{eq:Hamiltonian_real} describes both the
homogenous and trapped Bose gases.
For a homogenous gas, $\omega\_0 = 0$.
In both cases, we use a large enough $r\_b$ to minimize
the boundary effects.
We will set $\hbar = m = 1$ throughout this paper.


We discretize the cubic simulation box into an
$L\times L\times L$ lattice.
The lattice spacing is $\varsigma = 2 r_b / L$.
We enumerate the real-space sites using an integral index $i$
ranging from 1 through $L^3$.
The coordinate of the $i$-th site is given by $\rvec_i$.
The periodic boundary condition restricts the values for
the momentum coordinates
$\kvec = (k_1, ..., k_3)$
to
$k_i = \pi n_i / r_b$,
where
$n_i$ is an integer in the range
$\lfloor -L / 2\rfloor \le n_i < \lfloor L / 2\rfloor$.
We will use the index $q = 1, 2, ..., L^3$ to enumerate the points in the
momentum space; correspondingly, $\kvec_q$ is the momentum vector of the
$q$-th point.

The field operators on the lattice are defined to be
\eqsl{
    \label{eq:r-destroy}
    \Dc{\spin{\sigma}i}
&   \equiv
    \varsigma^{3/2} \Dpsi{\spin{\sigma}}(\rvec_i)
    \,,
\\
    \label{eq:q-destroy}
    \Db{\spin{\sigma}q}
&   \equiv
    \Dvarphi{\spin{\sigma}}(\kvec_q)
    \,.
}
The discretized Hamiltonian is therefore given by
\eql{eq:qHubbard}
{
    \Hop
& = \Half \spinsum{\sigma} \sum_{q}
    {\kvec}_q^2 \Cb{q\spin{\sigma}}\Db{q\spin{\sigma}}
  + \Half \left( \frac{\kappa}{\varsigma^2} \right)
    \spinsum{\sigma} \sum_{i}
    |\rvec\_i - \rvec\_0|^2
    \Cc{i\spin{\sigma}}\Dc{i\spin{\sigma}}
\\&
  +~\Half U \spinsum{\sigma} \sum_i
    \left(\Cc{i\spin{\sigma}} \Dc{i\spin{\sigma}}
          \Cc{i\spin{\sigma}} \Dc{i\spin{\sigma}}
          - \Cc{i\spin{\sigma}} \Dc{i\spin{\sigma}} \right)
    \,,
}
where
\eqsl{
    U
& = \frac{4\pi a_s}{\varsigma^3}      \label{eq:Hub2real-U}
    \,,
\\
    \kappa
& = \frac{\varsigma^2}{\aho[4]}        \label{eq:Hub2real-k}
    \,,
}
and $\aho \equiv \sqrt{\hbar / m \omega\_0}$ is the harmonic oscillator
length scale.
The representation of the kinetic energy in \Eq{eq:qHubbard}
reproduces the
continuum spectrum more faithfully than the real-space finite-difference
form in the original Hubbard form, and allows quicker convergence
with the size of the grid, $L$.

The contact two-body potential in the continuum is ill-defined
\Cite{Proukakis1998,Esry1999} because of the ultraviolet
divergence.
The momentum-space interaction strength,
\eq{
    \tilde{V}_{2B}(\mathbf{q})
    \equiv
    \int \!d\rvec\, V_{2B}(\rvec) e^{-i\mathbf{q}\cdot\rvec} \,,
}
is uniform for any $|\mathbf{q}|$.
The discretized Hamiltonian alleviates the problem to a large degree by
introducing a mometum space cut-off $k_c$ and replacing the
$\delta$-potential by an on-site interaction parameterized by the
scattering length, $a_s$.
%
%
However, the discretized two-body potential in \Eq{eq:qHubbard} must be
adjusted in order to yield the correct two-body scattering length,
and $a_s$ in \Eq{eq:Hub2real-U} must be replaced by an appropriate
$a_s'$ for the lattice.
Following the standard treatment,
we obtain the regularized $a_s'$, which for a 3D lattice is\Cite{Castin2004}
\eql{eq:REGa_s}
{
    a_s' \equiv
    \frac{a_s}
         {1 - 2.442749 a_s / \varsigma}
    \,.
}
For the system to be in the dilute limit and the form of our
two-body potential to be valid, we need the density at the lattice sites to
satisfy $\langle \hat n_i\rangle \ll 1$.

\section{Computational methods}
\label{sec:method}

\subsection{Quantum Monte Carlo method}
\label{ssec:QMC}

\subsubsection{General formalism for many-boson boson ground states}
\label{sssec:QMC-general}

We briefly describe our method of computing the ground state of
many bosons.
A detailed account can be found in \Ref{Purwanto2004}.
We project the ground-state wave function $\ket{\PsiGS}$ from a trial wave
function $\ket{\PsiT}$,
\eql{eq:g.s.proj}
{
     (\Pgs)^n_{} \ket{\PsiT}
     &\!\overset{n\rightarrow\infty}\longrightarrow\! \ket{\PsiGS}
     \,,
}
where $\ket{\PsiT}$ in this study is the GP solution (see
\Sec{ssec:GP} for details).
The projector
\eqsl{
    \Pgs
&   \equiv
    \expP{\Dt \ET} \expP{-\Dt\Hop}
    \label{eq:g.s.proj-def}
\\
&   =
    \expP{\Dt \ET} \expdtK \expdtV \expdtK
  + \Order(\Dt^2)
    \label{eq:g.s.proj-Trotter-def}
}
is evaluated stochastically by rewriting the two-body part
into a multidimensional integral.

The two-body part of the potential in \Eq{eq:qHubbard}
can be written as a sum of
the squares of one-body operators
$
    \Vop 
 = -\Half \sum_i \HSop_i^2
$,
where
$
    \HSop_i \equiv \sqrt{-U} \, \Cc{i}\Dc{i}
$
is essentially the density operator.
We use the following Gaussian integral identity to rewrite $\expdtV$ in
terms of the one-body operators:
\eql{eq:HS-xformns}
{
    \expP{\half\Dt \HSop^2}
= \frac{1}{\sqrt{2\pi}} \int_{-\infty}^{\infty} \! d\sigma \,
    \expP{-\half \sigma^2}
    \expP{\sigma \sigmabar - \half \sigmabar^2}
    \expP{\sqrt{\Dt} \, (\sigma - \sigmabar) \HSop}
    \,,
}
where the constant $\sigmabar$ is determined below.
We use an importance sampling scheme to sample the ground-state wave function,
so that
\eql{eq:MC-gswf}
{
    \ket{\PsiGS}
    \doteq
    \sum_{\{\wlkr{}{}\}}
    \Ovlp{\wlkr{}{}}{}
    \frac{\ket{\wlkr{}{}}}{\braket{\PsiT}{\wlkr{}{}}}
    \,,
}
where each $\ket{\wlkr{}{}}$ is a mean-field solution, i.e.,
a permanent consisting of identical single-particle orbitals.
In practice, this means that each $\ket{\wlkr{}{}}$ is
represented by a single-particle orbital.

The projection in \Eq{eq:g.s.proj}
is then realized by random walks in a manifold of mean-field
solutions\Cite{Zhang1997_CPMC,Purwanto2004},
which are governed by the following equation\Cite{Zhang2003,Purwanto2004}:
\eql{eq:Pgs-stochastic}
{
    \Wlkr[']{}{}
& = \int d\bm{\sigma} \, p(\bm{\sigma})
    \Bop(\bm{\sigma} - \bm{{\sigmabar}})
    W(\bm{\sigma}, \wlkr{}{})
    \Wlkr{}{}
    \,,
}
where
\eqsl{
    \label{eq:P-def}
    p(\bm{\sigma})
& = \prod_i \frac{\expP{-\half \sigma_i^2}}{\sqrt{2\pi}}
    \,,
\\
    \label{eq:B-def}
    \begin{split}
    \Bop(\bm{\sigma} - \bm{{\sigmabar}})
& = \expP{\Dt \ET}
    \expdtK
    \left\{
    \prod_i \expP{\sqrt{\Dt} \, (\sigma_i - \sigmabar[i]) \HSop_i}
    \right\}
\\
&   \times ~
    \expdtK
    \,,
    \end{split}
\\
    \label{eq:iz-prop-weight-hybrid}
    W(\bm{\sigma}, \wlkr{}{})
& =
    \frac{\ME{\PsiT}{\Bop(\bm{\sigma} - \bm{{\sigmabar}})}{\wlkr{}{}}}
         {\braket{\PsiT}{\wlkr{}{}}}
    \,
    \expP{\bm{\sigma} \cdot \bm{{\sigmabar}}
          - \half \bm{{\sigmabar}} \cdot \bm{{\sigmabar}}}
\,.
}
The optimal choice of the constant vector $\bm{{\sigmabar}}$
is\Cite{Zhang2003,Purwanto2004}:
\eql{eq:sigmabar-def}
{
    \sigmabar[i]
  = -\sqrt{\Dt} \,
    \frac{\ME{\PsiT}{\HSop\_i}{\wlkr{}{}}}
         {\braket{\PsiT}{\wlkr{}{}}}
    \equiv
    -\sqrt{\Dt} \, \bar{v}\_i
    \,.
}
With this choice, the weight factor in \Eq{eq:Pgs-stochastic}
can be written in the so-called local energy form\Cite{Zhang2003,Purwanto2004}:
\eql{eq:iz-prop-weight-EL}
{
    W(\bm{\sigma}, \wlkr{}{})
    \approx
    \expP{-\Dt \ME{\PsiT}{\Hop}{\wlkr{}{}} / \braket{\PsiT}{\wlkr{}{}}}
    \equiv
    \expP{-\Dt \EL(\wlkr{}{})}
    \,.
}
In practice, whether the local-energy or the hybrid form in
\Eq{eq:iz-prop-weight-hybrid} is more efficient will depend on the system.
For the calculations in this paper, we have mostly used the local-energy
form.

We initialize a population $\{\,\Wlkr{}{}\,\}$ to mean-field solutions,
e.g., $\ket{\PsiT}$.
A single random-walk step for each walker consists of updating
the orbital and its associated weight $\Ovlp{\wlkr{}{}}{}$,
\eqssl{eq:iz-prop}
{
    \label{eq:iz-prop-orbital}
    \Wlkr[']{}{}
&   \leftarrow
    \Bop(\bm{\sigma} - \bm{{\sigmabar}}) \Wlkr[]{}{}
\\
    \label{eq:iz-prop-weight}
    \Ovlp[]{\wlkr[']{}{}}{}
&   \leftarrow
    W(\bm{\sigma}, \wlkr{}{})
    \Ovlp[]{\wlkr{}{}}{}
    \,.
}
where the auxiliary fields $\{\sigma\_i\}$ are drawn from the Gaussian
probability density function $p(\bm{\sigma})$.

The computation of observables is done using the back-propagation
estimator\Cite{Zhang1997_CPMC,Purwanto2004},
\eql{eq:BP-obs}
{
    \BPEst{\Aop}
  = \frac{\ME{\PsiT}{\exptBPH\Aop}{\PsiGS}}
         {\ME{\PsiT}{\exptBPH}{\PsiGS}}
    \,,
}
which for large enough $\tauBP$ yields the ground-state expectation
value for any observable.

\subsubsection{Phaseless approximation}
\label{sssec:QMC-phaseless}

The formalism above is exact. For repulsive interactions,
unfortunately, $\HSop_i$ in \Eq{eq:HS-xformns} becomes imaginary.
This is similar to the phase problem in fermionic systems\Cite{Zhang2003},
and we apply the recently developed
\emph{phaseless approximation\/}, which has been shown to work
well in electronic-structure calculations\Cite{Zhang2003}.
This method eliminates the phase problem at the cost of
a systematic bias which is dependent on the trial wave function.
As we will demonstrate in benchmark calculations in
Appendix~\ref{app:QMC-phaseless-bench},
the bias is relatively small for the bosonic systems we study here.
Indeed, for all but the largest values of
$a_s$, it is possible to perform unconstrained calculations
with fixed imaginary-time, $\beta=n\Dt$, in which $\beta$ can be made
sufficiently long that essentially exact ground-state values are obtained.
Comparison with these results shows that the systematic error in the
phaseless approximation is small (see
Appendix~\ref{app:QMC-phaseless-bench}).

In the phaseless approximation, the weights $\{w_{\wlkr{}{}}\}$ are restricted
to real, positive values.
We define the phase rotation angle $\Delta\theta$ by
\eql{eq:dtheta-def}
{
    \Delta\theta
    \equiv
    \Im \ln \left(\frac{\braket{\PsiT}{\wlkr[']{}{}}}
                       {\braket{\PsiT}{\wlkr{}{}}}
            \right)
    \,.
}
This is the complex-phase rotation of the walker's overlap with
the trial wave function as a result of the application of
$\Bop(\bm{\sigma} - \bm{{\sigmabar}})$ to $\Wlkr{}{}$.
In the phaseless approximation, the evolution
of $\Ovlp[]{\wlkr{}{}}{}$ is altered to
\eql{eq:ph-iz-prop-weight}
{
    \Ovlp[]{\wlkr[']{}{}}{}
    \leftarrow
    \begin{cases}
      \cos(\Delta\theta) |W(\bm{\sigma}, \wlkr{}{})| \Ovlp[]{\wlkr{}{}}{}
      \,,
      & |\Delta\theta| < \pi/2
      \\
      0
      \,,
      & \text{otherwise}
    \end{cases}
    \,,
}
which prevents the walkers from reaching the
origin of the $\braket{\PsiT}{\wlkr[]{}{}}$-complex-plane.
Equations \eqref{eq:iz-prop-orbital} and \eqref{eq:ph-iz-prop-weight}
define the algorithm of the phaseless QMC method.

In invoking the phaseless approximation, it is helpful to rearrange
the two-body interaction term in $\Hop$ such that a mean-field background
is subtracted:
\eql{eq:shiftV2}
{ \Vop = -\Half \sum_i (\HSop_i - \langle \HSop_i \rangle)^2
-\sum_i \HSop_i \langle \HSop_i \rangle
+ \Half \sum_i \langle \HSop_i \rangle^2
\,,
}
where the constant $\langle \HSop_i \rangle $ is the mean-field
expectation value, e.g., with respect to $\ket{\PsiT}$.
The residual term involving $(\HSop_i - \langle \HSop_i \rangle)$ is then
used in \Eq{eq:HS-xformns}.
This would have \emph{no effect} in the exact formalism above, where, as
we discussed in \Ref{Purwanto2004}, the importance sampling transformation
effectively introduces the background subtraction even if the bare form
of $\Vop$ is used.
With the phaseless approximation, however, the rotation angle
is controlled by the mixed-estimate of $\HSop_i$.
Reducing its average by subtracting the mean-field background will thus
help reduce the rotation, and improve the behavior of the approximation
in \Eq{eq:ph-iz-prop-weight}.

We note that the presence of phaseless approximation breaks the
time-reversal symmetry of the ground-state projector.
The forward, phaseless propagator $(\exptBPH)_{\mathrm{ph}}$ is
no longer formally equivalent to the back-propagated, phaseless propagator
$(\exptBPH)^{\dagger}_{\mathrm{ph}}$ [see \Eq{eq:BP-obs}].
This results in an additional systematic error in the back-propagation
estimator.
The expectation value of an operator $\Aop$ computed from back-propagation is
$
    \ME{\PsiGS''}{\Aop}{\PsiGS'}
$,
where $\ket{\PsiGS'}$ and
$\ket{\PsiGS''}$ are the approximate ground-state wave functions
(normalized) in the forward- and backward-direction, respectively, and
they are in general not the same.
This is similarly the case in the constrained-path Monte Carlo for fermion
lattice models \Cite{Zhang1997_CPMC,
Carlson1999}.
It was shown\Cite{Carlson1999} that the error vanishes linearly as
$\ket{\PsiT} \rightarrow \ket{\PsiGS}$.
We will further discuss the effect of the phaseless constraint in
\Sec{sec:Discussions} and Appendix~\ref{app:QMC-phaseless-bench}.

\subsection{GP self-consistent projection and QMC trial wave functions}
\label{ssec:GP}

We solve the GP equation on the same lattice defined for QMC, using a
self-consistent projection with the GP propagator
$\exp{(-\Dt \HGP)}$\Cite{Purwanto2004}.
Aside from a factor $(N-1)/N$ in front of the interaction terms,
the one-body Hamiltonian $\HGP$ is simply \Eq{eq:qHubbard} with the
replacement
\eql{eq:op2b_GP}
{
    \Cc{i} \Cc{i} \Dc{i} \Dc{i}
    \rightarrow
    2 \XV{\Cc{i}\Dc{i}} \, \Cc{i}\Dc{i} -
    \XV{\Cc{i}\Dc{i}}^2
    \,,
}
where the expectation is with respect to the GP solution.
As discussed in \Ref{Purwanto2004}, our QMC can be thought of as
stochastically carrying out the functional integral, while GP is
the saddle-point approximation.

The $U$ parameter in the GP calculations is given by
the bare $a_s$ rather than the
regularized $a_s'$ using \Eq{eq:REGa_s},
because the shape-independent $\delta$ potential has become a mean-field
potential in the GP approximation.
It is these GP results that we compare with.


For our QMC calculations, the trial wave function $\PsiT$ is taken to be
the solution of the GP-like projection, but with the regularized $a_s'$.
This wave function is different from the correct GP solution above,
which is obtained using the bare $a_s$.
Each value of the discretization parameter $\varsigma$ corresponds to a
different renormalized $a_s'$ [see \Eq{eq:REGa_s}], and gives rise to
distinctly different results, while the correct GP solution converges
rapidly with $\varsigma$ (see \Fig{fig:coldens-as400.100p.fzfx4}).
As the trial wave function, however, we argue that the optimal choice is
the best variational solution, which is given by the corresponding
mean-field calculation with the same $a_s'$.

\subsection{Bogoliubov approximation}  
\label{ssec:Bogoliubov}

In the Bogoliubov approximation\Cite{Bogoliubov1947,Lee1957,Wu1959},
correlation effects are treated by means of perturbation,
where the zeroth-order term is the GP mean-field solution.
The approach was first formulated for a homogenous Bose gas.
It assumes a macroscopic occupancy of the lowest
energy state ($\kvec = \mathbf{0}$), namely $(N - N_0) \ll N$.
For each density $\rho = N / \Omega$ and interaction strength,
the total energy per particle $\EBog/N$, momentum distribution
$\pi_{\mathrm{Bog}}(\kvec)$, and condensate fraction $N_0/N$
can be written down analytically
in the thermodynamic limit.
The corrections to the mean-field GP results are expressed in terms of
the gas parameter $\rho a_s^3$, which gives a measure of the
deviation from the mean-field picture.
Note that the bare $\as$ should be used, since
the regularization of the scattering length is implicitly done in
the Bogoliubov approximation as is in GP.

It is important to truncate the summation over
${\mathbf k}$ in computing the momentum distribution and kinetic energy.
This stems from the incorrect behavior of the Bogoliubov $\pi(\kvec)$
at large momenta:
$
    \pi\_{\mathrm{Bog}}(\kvec)
    \propto
    1/|\kvec|^4
$
as
$|\kvec| \rightarrow \infty$.
Physically,
the form of the two-body potential requires that $|\kvec| \as \ll 1$,
therefore the contribution from $|\kvec|$ larger than
a cutoff momentum $k_c$ should be excluded.
We use an explicit numerical summation with the same $\kvec$-space grid as
in QMC.
This automatically limits the sum to the reciprocal lattice
(excluding $\kvec = \mathbf{0}$).
In addition, it helps to correlate the finite-size
effects in the two calculations, and allows for
a more direct comparison of the results between Bogoliubov and QMC.

We extend the Bogoliubov approach to the inhomogeneous case using a
local-density approximation (LDA), by treating each
lattice site as a locally homogenous Bose gas.
This is similar to the LDA approximation for electronic
systems under density functional theory\Cite{Kohn1999},
and we refer to it as \emph{Bogoliubov-LDA}.
The approximation is expected to be reasonable
if the density is smooth and slowly varying, which
is fulfilled in our dilute Bose gas systems.

The kinetic energy, for example,
is a sum of two contributions under this approach: one from the
curvature (inhomogeneity) of the density profile, and the other from
Bogoliubov correction.
Given the real-space density $\rho(\rvec)$, it is
\eql{eq:LDABog-KE}
{
    \XV{\Top}\_{\mathrm{Bog}\textrm{-}\mathrm{LDA}}
&=  -\Half \int d^3\rvec \,
    \sqrt{\rho(\rvec)} \, \nabla^2 \sqrt{\rho(\rvec)}
\\
&+~ 
    \int d^3\rvec \,
    \widetilde{T}_{\mathrm{Bog}}[\rho(\rvec)]\rho(\rvec)
    \,,
}
where the functional $\widetilde{T}_{\mathrm{Bog}}[\rho(\rvec)]$
is the Bogoliubov kinetic energy per particle for a gas with uniform
density $\rho = \rho(\rvec)$.
More details on our Bogoliubov-LDA procedure
are provided in Appendix~\ref{app:Bog-gs}.)

\section{Results}
\label{sec:Results}

In this section, we present results on
the energetics, condensate fraction,
density profile, and momentum distribution.
Individual energy terms are computed: $\kinXV$ is the kinetic energy,
$\potXV$ the two-body interaction energy,
and $\trapXV$ the external trapping potential.
\COMMENTED{
We simulate systems containing up to $1000$ atoms.
We expect to be able to deduce the physical behavior of the realistic
condensate, having much larger number of particles, from our observations
here.
}

In the calculations,
we typically use a $24 \times 24 \times 24$ lattice,
with a simulation box of
linear dimension $2r\_b = 14 \aho$.
This gives us a lattice constant of $\varsigma = 0.583 \aho$.
Our trap length is $\aho = 8546\,\Angstrom$, which gives typical
peak densities of about $ 10$ to $40\,\um^{-3}$ for $100$ to $1000$
particles in the trap.
The lattice constant $\varsigma$ is large compared to our
scattering lengths (up to
$a_s \sim 1000\,\Angstrom$), which is consistent with the assumption
in neglecting the details of the two-body potential.

%
\begin{figure}[htbp]

  \includegraphics[scale=1.3]{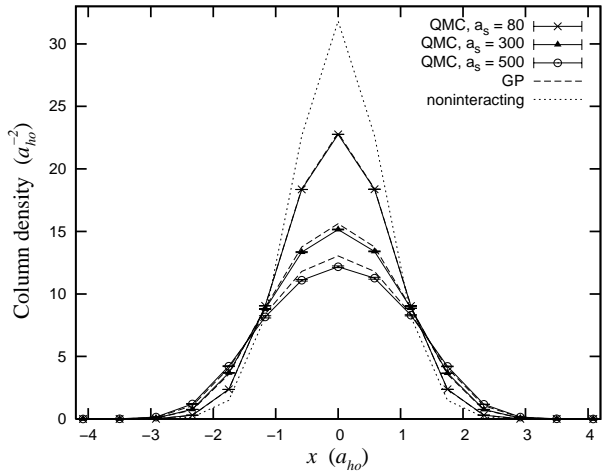}
  \caption{\label{fig:coldens-scan-as.100p.D=14}%
The ground-state column density $\rho_{y}(x, z=0)$ of a trapped gas
containing $N = 100$ bosons with three different scattering lengths:
$a_s =80\,\Angstrom$, $300\,\Angstrom$, and $500\,\Angstrom$.
QMC statistical error bars are indicated.
The GP densities are next to the corresponding QMC curve, and are all shown
in dashed lines.
Also shown as a reference is the non-interacting profile.
  }

\end{figure}

\subsection{Density Profile} 
\label{ssec:Results_density}

\begin{figure}[!hbtp]

  \includegraphics[scale=1.4]{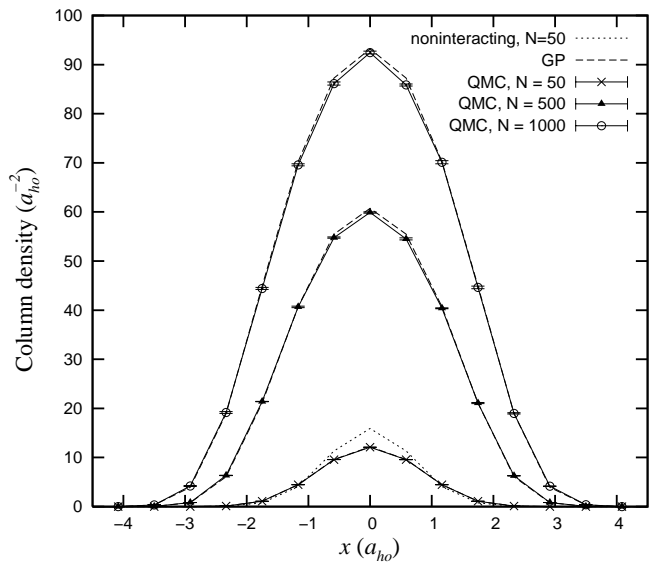}
  \caption{\label{fig:coldens-scan-Np.as120.D=14}%
The ground-state column density $\rho_{y}(x, z=0)$ of a trapped gas of
$N = 50$--$1000$ bosons with scattering length of $a_s = 120\,\Angstrom$.
  }

\end{figure}

Figure~\ref{fig:coldens-scan-as.100p.D=14} shows the density profiles
of 100 trapped bosons for three different scattering lengths.
To make a connection with experiments, we show the column density
\eql{eq:col-dens-def}
{
    \rho_{y}(x, z)
    \equiv
    \int \!dy\, \rho(x,y,z)
    \,,
}
that is, the density integrated along a particular direction
(e.g., the $y$-axis),
which can be observed through optical
measurements\Cite{Andrews1996,Andrews1997,Hau1998}.
As we increase $a_s$,
the condensate expands due to the increasing repulsive interactions.
Similarly, as we add more particles into the gas,
the density profiles expands,
as shown in \Fig{fig:coldens-scan-Np.as120.D=14}.

Compared to GP, the QMC peak density is always lowered, and
the QMC overall density profile is more extended.
For $a_s = 80\,\Angstrom$, the peak column density is lowered by $0.5\%$
from GP.
For $a_s = 500\,\Angstrom$, this difference is about $7\%$.
Earlier many-body calculations using
the correlated basis approach\Cite{Fabrocini1999,Fabrocini2001} and
DMC\Cite{Blume2001,DuBois2003} also showed the same qualitative behavior.
Below we will further discuss these in connection with the energetics
and momentum distribution.

\begin{figure}[!hbtp]

  \includegraphics[scale=1.5]{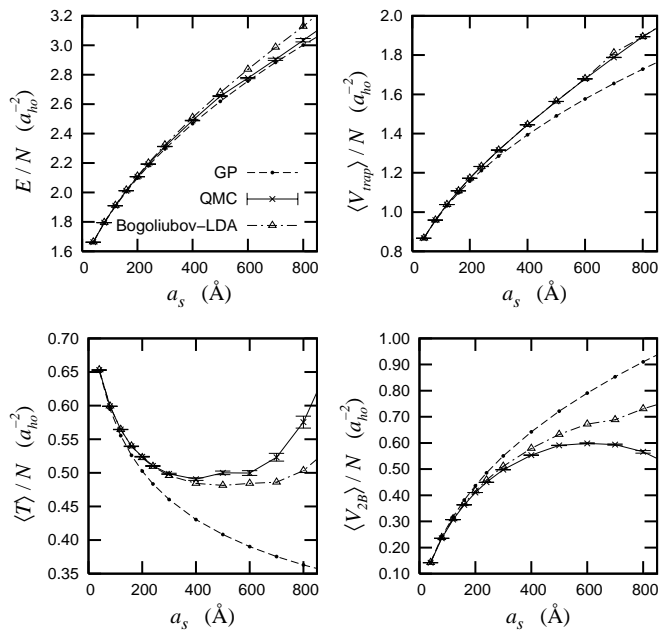}
  \caption{\label{fig:scan-as.D=14}%
Ground-state energy per particle and its individual components as a function
of the scattering length. The system has
$100$ bosons. QMC error bars are statistical.
  }

\end{figure}

\subsection{Energetics}  
\label{ssec:Results_energy}

Figure~\ref{fig:scan-as.D=14} shows the ground-state energy
and its individual components as a function of the
interaction strength. We see that,
as the scattering length $a_s$ is increased, the total energy
increases as expected.
Both GP and Bogoliubov-LDA energies are in reasonable agreement with
QMC, deviating more at larger $a_s$. The GP energy is slightly lower
than the exact results (no variational principle due to regularization),
while Bogoliubov-LDA is higher.
The external potential energy, $\trapXV$,
also increases with $a_s$, which is a consequence of the expansion of the
density profile with interaction,
as shown in \Fig{fig:coldens-scan-as.100p.D=14}.
The GP trap energy is lower than QMC, consistent with the result in
\Fig{fig:coldens-scan-as.100p.D=14} that
QMC density profiles are more extended.





The kinetic and interaction energies are shown in the bottom panels of
\Fig{fig:scan-as.D=14}.
The discrepancy between GP and QMC is more pronounced.
In particular, the GP kinetic energy decreases
monotonically with $a_s$, because the density profile expands and the
system becomes less confined.
The QMC kinetic energy, on the other hand,
shows a \emph{nonmonotonic behavior}.
For small $\as$, the kinetic energy decreases as $\as$ is increased,
tracking the GP result. At $\as \gtrsim 400\,\Angstrom$, however,
the kinetic energy curves up and increases with $\as$.
The QMC interaction energy is significantly lower than the mean-field
interaction energy at large $a_s$, and
the GP result increases much more rapidly with $\as$ than
QMC. Indeed the QMC curve appears to turn downward at the last point,
but our data is not sufficient to establish this, as it is possible that
a larger systematic error from the
phaseless approximation may have contributed to make the QMC result
smaller
(see the benchmark results in Appendix~\ref{app:QMC-phaseless-bench}).

\begin{figure}[!thbp]

  \includegraphics[scale=1.4]{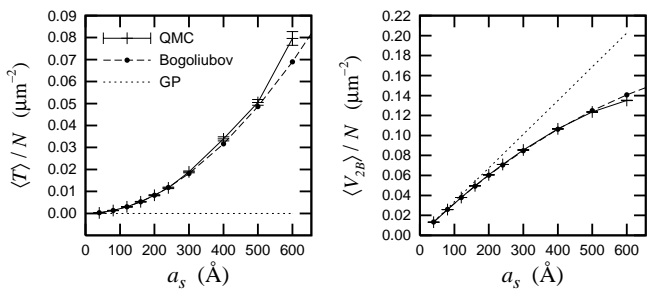}
  \caption{\label{fig:Uscan-as.100p.D=6.66-K+V2b}%
Kinetic and interaction energies in the \emph{uniform Bose gas}
as a function of the scattering length.
We show results from QMC, Bogoliubov, and GP.
The density is $\rho = 0.542 \,\um^{-3}$.
The simulation box has 100 particles on a $16 \times 16 \times 16$ lattice,
representing a physical volume of $\Omega = 184.4\,\um^3$.
}

\end{figure}

From a single-particle picture, we would expect the QMC kinetic energy
to be lower than that of GP, since the QMC density profiles are more extended.
In reality, correlation effects become more important as $a_s$ increases,
which raises the kinetic energy with interaction.
This is illustrated clearly by considering the uniform Bose gas,
for which we show corresponding results in
Fig.~\ref{fig:Uscan-as.100p.D=6.66-K+V2b}.
The GP ground state is a zero-momentum condensate.
In the many-body ground state,
interactions excite particles into
higher-momentum single-particle states,
raising the kinetic energy as a result.
The QMC results in the trapped gas are thus the outcome of
the competition between mean-field and correlation effects.

The Bogoliubov-LDA calculations,
whose results are also shown in \Fig{fig:scan-as.D=14},
help to quantify this picture further.
We use QMC density profiles in the calculation
(hence the exact agreement between the
Bogoliubov-LDA and QMC estimates of the trap energy in \Fig{fig:scan-as.D=14}),
although we have verified that
the physics is qualitatively unchanged if the GP
densities are used instead.
The result shows good agreement with the full many-body calculation.
In particular, the Bogoliubov kinetic energy shows an increase
similar to the QMC prediction.
The corresponding interaction energy is also reduced, although not as much
as in QMC.
Overall, the Bogoliubov results capture the basic picture and confirm that
correlations are an important ingredient in the energetics of the gas.

\subsection{Condensate fraction and momentum distribution} 
\label{ssec:pdensity}

\begin{figure}[!tbhp]

  \includegraphics[scale=1.4]{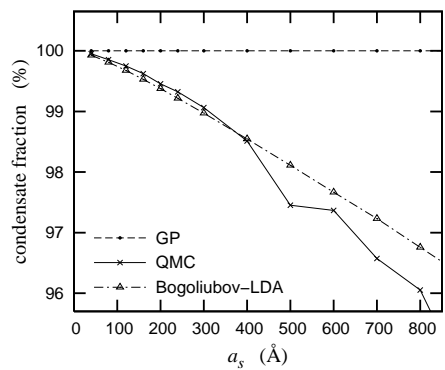}
  \caption{\label{fig:cond-frac.100p.D=14}%
Condensate fraction as a function of the scattering length.
The system is the same as that of \Fig{fig:scan-as.D=14}.
The condensate fraction is defined as the
leading eigenvalue (normalized by $N$)
of the one-body real-space density matrix.
QMC statistical errors were not shown.
  }

\end{figure}

\begin{figure}[hbtp]
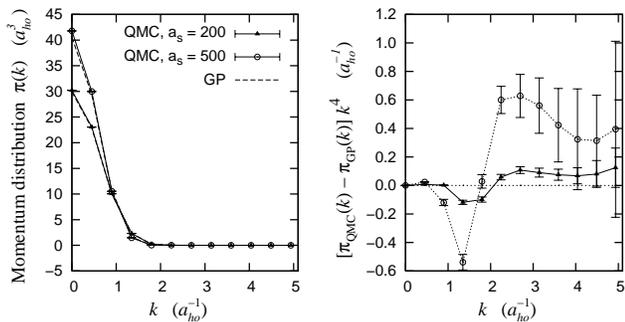


  \includegraphics[scale=0.96]{\PLOTFILE{T3D+24s100p.pdensity.eps}}
  \includegraphics[scale=0.96]{\PLOTFILE{T3D+24s100p.pdensity-delta.k4.eps}}

  \caption{\label{fig:pdensity.100p.D=14}
Momentum distribution for trapped gases at two different scattering lengths.
The system again has $N = 100$ bosons.
The left panel shown a cut along the $k$-axis.
The right panel shows the difference between the QMC and GP,
multiplied by $k^4$.
  }

\end{figure}

Figure~\ref{fig:cond-frac.100p.D=14} shows the condensate fraction
as a function of interaction strength. GP by definition gives 100\%. We
see that the actual depletion is about $4\%$ at $800\,\Angstrom$.
Again, the Bogoliubov result agrees well with QMC.
Figure~\ref{fig:pdensity.100p.D=14} shows the momentum
distribution for two scattering lengths:
$a_s = 200\,\Angstrom$ and $500\,\Angstrom$.
The QMC's momentum distribution is more peaked than GP.
This translates in the real space to a more extended density profile
for QMC, as is observed in \Fig{fig:coldens-scan-as.100p.D=14}.

The graph also shows the contribution to the kinetic energy from various
$k\equiv |\kvec|$ regions, since the kinetic energy is related to the
momentum distribution through
\eql{eq:KE-pdensity}
{
    \kinXV
&   \propto \int k^2 dk \, \pi(k) k^2
    \,.
}
Relative to GP, the QMC distribution is depleted in the medium-$k$
regime, around $k \sim \aho[-1]$.
Part of this depletion goes to the low-momentum region near $k = 0$,
and the other to the high-$k$ region.
At a higher $a_s$, the depletion shifts toward the smaller $k$ region.
It is clear that the enhancement in the high-$k$ region results in
the increase of the kinetic energy.
The kinetic energy is strongly enhanced in the larger $\as$ cases, which
results in the upturn of the kinetic energy curve in
\Fig{fig:scan-as.D=14}.

A precision measurement of the momentum distribution would be useful to reveal
the detailed structure of the many-body correlations in the Bose gas.
Our results from a lattice do not have enough resolution to
reveal whether there are finer structures in the momentum- or
real-space density.
(A fine structure in the density profile was predicted by the DMC
calculations\Cite{DuBois2003}.)
In the auxiliary-field QMC framework, a better resolution in the density
profile may be obtained by choosing a more suitable basis set,
such as
Hartree-Fock states\Cite{Esry1997},
whereby the GP solution becomes the lowest-energy state in this basis set,
and also the leading solution in the ground-state wave function.

\section{Discussions}
\label{sec:Discussions}

\subsection{GP, modified GP, and Bogoliubov-LDA approaches}
\label{sec:Bog-vs-modGP}

We have shown that the many-body correlations qualitatively change the
behavior of the kinetic energy in the trapped Bose gas.
The Bogoliubov approximation\Cite{Bogoliubov1947, Lee1957, Wu1959}
under the local density approximation (LDA), which we refer to as
Bogoliubov-LDA, captures this trend quite well.
The LDA provides a good way to include the correlation effects based
on the homogenous Bose gas results.
This is perhaps not surprising, given the diluteness of the gas.

In contrast, the mean-field GP method
by construction approximates the kinetic energy only by the part that
arises from the inhomogeneity of the gas, missing the portion from
many-body effects.
The separation of these two portions is especially clear in the
homogeneous gas, as we illustrated in \Sec{ssec:Results_energy}.
This appears to be a rather generic feature of independent-particle approaches.
The same would apply to the modified GP (MGP)
method\Cite{Braaten1997, Nunes1999, Fabrocini1999, Banerjee2001,
Fu2003}, which can be viewed
as the bosonic counterpart of the standard electronic structure method of
LDA under density-functional theory (DFT).
In that framework, the MGP equation is an outcome of using the Bogoliubov
results for the uniform Bose gas as the ``exchange-correlation'' (xc)
functional, i.e., LDA+Bogoliubov ({\emph{as opposed to} the
Bogoliubov-LDA above).
This method has a great advantage in that it allows self-consistent
calculations.
For example, the real-space density can be calculated directly
and would not need to be imported as was done with the Bogoliubov-LDA.
Further, it is of course possible to use exact QMC results on the uniform gas
to further improve the MGP equation, and make it more like DFT-LDA.
For the kinetic energy, however, the MGP would give the same qualitative
results as GP, even when the exact xc-functional is used and
the exact density is obtained, because the ``kinetic energy'' that is
explicitly defined in the MGP framework is incomplete.
In fact, the same would seem to apply to DFT-LDA for electronic systems.
This is an important conceptual difference between MGP and Bogoliubov-LDA
approach, although they are closely related and lead to the same total
energy results.

\subsection{Finite-size effects and limitations of the on-site potential} 
\label{ssec:pot-details}

There are two kinds of finite-size errors in our calculation: the error
due to finite simulation box size, and the discretization error due to
finite lattice constant.
The first kind is easily reduced, by
increasing the simulation box size, $r_b$.
In the trapped boson calculations with $N=100$ particles, we have checked
that $r\_b \gtrsim 5 \aho$ is sufficient for $a_s \le 1000$.
For calculations with large values of $N$, we use
$r\_b = 7 \aho$ to allow simulations of large enough condensate while
keeping the finite-size errors much less than our statistical error.

\COMMENTED{
\begin{figure}[!htbp]

  \includegraphics[scale=1.4]{\PLOTFILE{T3D+100p.as500.fzfx1.eps}}
  \caption{\label{fig:fzfx1-rb}%
The effect of finite simulation box size ($r\_b$).
The test system is a 100-particle gas with $\aho = 8546\,\Angstrom$.
The lattice constant is $\varsigma = 0.583\aho$.
On this scale, the energies are converged for $2 r\_b / \aho > 8$.
  }

\end{figure}
}

The discretization error from the finite lattice constant, $\varsigma$, is
more subtle.
On the one hand, sufficiently small $\varsigma$ should be used
so the results converge to the continuum values.
Figure~\ref{fig:fzfx2-zeta} shows the convergence of the total energy.
It also illustrates the effect of regularizing the
scattering length, as discussed in \Sec{sec:model'}.
In \Fig{fig:coldens-as400.100p.fzfx4}, we show the convergence of
the density profile.
\begin{figure}[!htbp]

  \includegraphics[scale=1.4]{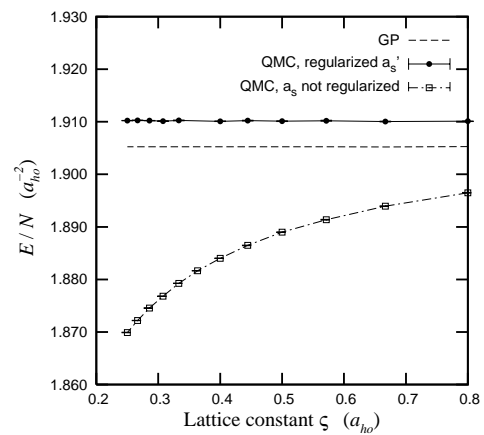}
  \caption{\label{fig:fzfx2-zeta}%
The effect of finite discretization on the QMC and GP total energies
due to the lattice constant $\varsigma$.
The test system has $N = 100$,
$a_s = 120\,\Angstrom$.
We show the total energy of the system for $\varsigma$ ranging
from $0.8\,\aho$ ($10 \times 10 \times 10$ lattice)
through $0.25\,\aho$ ($32 \times 32 \times 32$ lattice).
Also shown is the QMC energy calculated \emph{without} regularizing 
the scattering length, which
fails to converge. 
The statistical error bars are smaller than the point size.
  }

\end{figure}
%
%
\begin{figure}[!htbp]

  \includegraphics[scale=1.4]{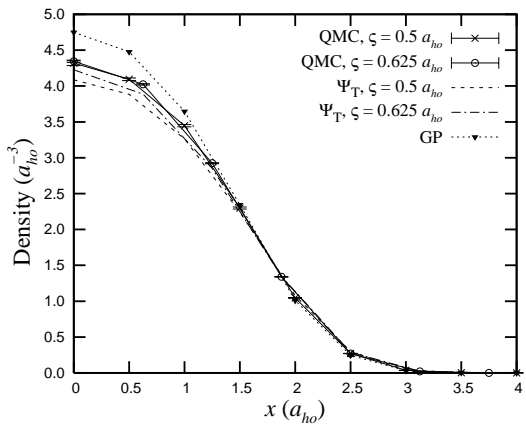}
  \caption{\label{fig:coldens-as400.100p.fzfx4}%
The effect of finite discretization on the density profiles in QMC and GP.
The density profile $\rho(x, y=0, z=0)$ is shown for
two different values of the lattice constant, $\varsigma$.
The test system has $N=100$ and $a_s = 400\,\Angstrom$.
The GP curves are indistinguishable. The densities
obtained from the QMC trial wave functions are also shown.
These are GP-like solutions but with regularized scattering lengths.
They do not converge like the QMC or true GP densities.
  }

\end{figure}

On the other hand, the lattice constant is also coupled to the on-site
potential that we use, which in turn affects the detailed energetics of
the system.
The on-site potential effectively has finite range and strength which
depend on $\varsigma$.
This is equivalent to setting the cutoff momentum $k_c \propto 1 /
\varsigma$ in the interaction matrix elements.
Figure~\ref{fig:fzfx-zeta.scan-as} shows the total and kinetic
energies as $\varsigma$ is varied.
The total energy is less sensitive to the details of the interaction potential,
as are the real-space density
(see \Fig{fig:coldens-as400.100p.fzfx4})
and the trap energy.
The dependence on $\varsigma$ in the kinetic and interaction energies,
however, is not negligible.
(This dependence is consistent with the observation of Mazzanti and
co-workers\Cite{Mazzanti2003} when they
varied the range of their soft-sphere repulsive potential.)
It is important to note that the nonmonotonic behavior
of the kinetic energy is observed at all $\varsigma$ values.
As $\varsigma$ is reduced, the upturn is more enhanced,
indicating a stronger effect from the interactions as the potential is made
narrower and harder.

\begin{figure}[!htbp]

  \includegraphics[scale=1.4]{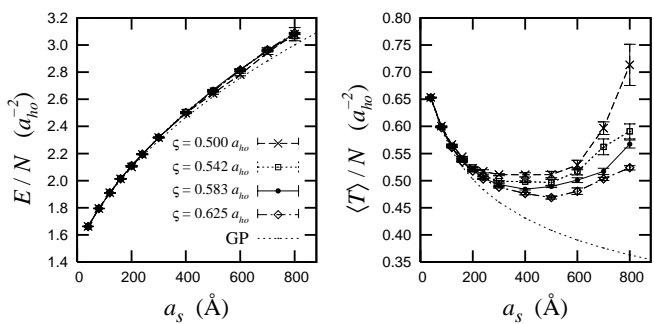}
  \caption{\label{fig:fzfx-zeta.scan-as}%
The effect of the lattice discretization and on-site interaction.
The total energy is insensitive to $\varsigma$,
whereas the kinetic energy shows more dependence on the details of the
interaction.
The system is the same as in Fig.~\ref{fig:scan-as.D=14}.
  }

\end{figure}


Ideally, we would like to decouple the basis-size error (due to
finite lattice spacing) from the effect of the details of the potential.
For this purpose, the on-site pseudopotential is inadequate.
The $\delta$-function potential is meant to be used with
the short-distance contributions already ``integrated out''\Cite{Leggett01}.
The effects above represent corrections from the details of the interaction
potential as defined by the on-site form, which change as we vary $\varsigma$.
It is easy to see that in the limit of $\varsigma \rightarrow 0$, the gas
is trivially noninteracting in the exact many-body
picture\Cite{Braaten1997}, since the range of the interaction potential
is zero.
However, if the conditions for the validity of the potential are maintained,
the corrections should be small and not affect essential properties,
as we have illustrated.
A better pseudopotential should have an intrinsic decay in momentum space
with well-defined convergence properties.

\subsection{Bias due to phaseless approximation}

The phaseless approximation, as demonstrated by the benchmarks in
Appendix~\ref{app:QMC-phaseless-bench},
gives an excellent approximation to the true
many-body ground state for weak to moderate interaction strengths.
Nevertheless, systematic errors on the computed observables are expected.
For example, the variational principle, that the total energy computed by QMC
is an upper bound to the exact energy, is not
guaranteed in the presence of phaseless approximation\Cite{Zhang2003,Carlson1999}.
We even observe this bias in the $a_s = 500\,\Angstrom$ results shown in
Table~\ref{tbl:T3D+phls-benchmark-as80-500}.

The systematic bias is noticeable, but
remains quite small up to the largest scattering lengths we study,
as can be seen from the benchmark data.
It is interesting to compare the phaseless and unconstrained
QMC energies in Table~\ref{tbl:T3D+phls-benchmark-as80-500}.
At a large $a_s = 500\,\Angstrom$, the phaseless approximation lowers the
kinetic energy (as well as the interaction energy) compared to the
unconstrained result.
This trend is observed for all $a_s$ values.
Since the phaseless bias increases with the interaction strength, it
should lead to an \emph{underestimation} of the upturn of
the kinetic energy.
Thus the nonmonotonic behavior of the kinetic energy should actually be
slightly stronger than shown by QMC.


We have shown in \Ref{Purwanto2004} that the QMC results is independent
of the input trial wave function $\PsiT$.
This is no longer the case in the presence of the phaseless approximation.
The approximation imposes a
constraint based on the overlap $\braket{\PsiT}{\wlkr{}{}}$,
and each $\PsiT$ in principle has different constraining properties.
This dependence is very weak, however, as we observed in our calculations
among trial wave functions of the same general form (GP-like).

The phaseless approximation can also affect the Trotter error,
which arises from the use of a finite time step $\Dt$ in
\Eq{eq:g.s.proj-Trotter-def}.
This error is controllable, and can be
extrapolated away by running at different values of $\Dt$.
Because the rotation angle in the random walk
is proportional to $\sqrt{\Dt\,U}$, the severity of the phaseless
projection is affected by $\Dt$, as is the extent of the population
fluctuation.
The latter is important in back-propagation, where it is highly desirable
to keep branching to a minimum.
If phaseless projection causes significant loss of the population, the
Trotter error will be increased.
Procedures that reduce the extent of the phase projection, for example,
subtracting the mean-field background shown in \Eq{eq:shiftV2},
will thus improve computational efficiency (in addition to possibly
reducing the systematic error).

\section{Conclusions}
\label{sec:Conclusions}

We have studied the ground state of realistic systems of trapped
interacting Bose atomic gases
using a many-body auxiliary-field QMC method, as well as GP and the
Bogoliubov method under a local density approximation.
We observed the effect of correlations in the energetics, condensate fraction,
real-space density profiles, and momentum distribution.
The density profile is more expanded compared to the GP prediction.
The momentum distribution shows enhancement in the occupation of the low-
and high-momentum states.
The kinetic energy, contrary to the GP estimate, is \emph{not} monotonic
with the scattering length $\as$.
The Bogoliubov method is able to reproduce this trend qualitatively.
Additional calculations on the uniform Bose gas were performed to
help understand and quantify our results.

Through this study we also further tested and developed our QMC method.
We found that the phaseless approximation developed for electronic
systems\Cite{Zhang2003} worked quite well in the context
of boson calculations with repulsive calculations.
Because of the simplicity of these bosonic systems compared to electronic
systems, they have provided an ideal testbed and allowed us to carry out
more benchmark calculations and gain additional insights on controlling
the phase problem, which is crucial for making QMC more useful for a wide
variety of problems.
It is hoped that the formalism we developed
will allow the study of many interacting Bose,
Fermi, and mixed-species systems.
The method can also account for different external experiment environments
(1-D or 2-D, rotations, anisotropic traps, optical lattices, etc.) quite
straightforwardly.

\begin{acknowledgments}

It is a pleasure to thank Markus Holzmann and Henry Krakauer
for stimulating discussions.
We gratefully acknowledge financial support from NSF and ONR.
We also thank the College of William and Mary's Computational
Science cluster (SciClone) project and the Center of Piezoelectric by Design for
computing support.

\end{acknowledgments}

\appendix

\section{Benchmark results on the phaseless approximation in QMC}
\label{app:QMC-phaseless-bench}

In this appendix, we show benchmark results on
the phaseless approximation in dealing the complex-phase problem,
as discussed in \Sec{sssec:QMC-phaseless}.
We first show results on a small system for which exact diagonalization
can be done.
We choose a one-dimensional Bose-Hubbard system.
The corresponding Gross-Pitaevskii calculation is also done at the same
Hubbard parameters $t$, $U$, and $\kappa$.
(Here $U$ is a fixed parameter which is the same in QMC and GP.)
Table~\ref{tbl:T1D+13s4p.phls-benchmark} compares the energetics and
condensate fraction obtained using various methods: exact diagonalization,
our QMC with the phaseless approximation (ph-QMC), and
GP self-consistent projection.

The ph-QMC improves over GP, and in general agrees well with exact
diagonalization.
The bias due to the phaseless approximation
is visible in the trap energy $\trapXV$.
In our phaseless QMC calculation, the mean-field background was
subtracted in the Hamiltonian, as shown in \Eq{eq:shiftV2}.
Applying the phaseless approximation directly leads to more population
fluctuations in the random walk and larger systematic errors in
$\trapXV$ and $\potXV$.
\begin{table}[!htbp]
\begin{center}

  \caption{\label{tbl:T1D+13s4p.phls-benchmark}
Benchmark of QMC with the phaseless approximation
(ph-QMC) against exact diagonalization.
The test system is a 1D Bose-Hubbard Hamiltonian with $13$ sites and
$4$ particles.
The parameters are $t = 2.676$, $U = 1.538$, and $\kappa = 0.3503$.
QMC statistical errors are in the last digit, and are shown in
parantheses.
Error bar in the condensate fraction was not estimated.
  }
  \begin{tabular}{lllllr}
  \hline
  \hline
   Type           &  $E$          & $\kinXV$     &  $\trapXV$   & $\potXV$    &  $(N_0/N)$ \\
  \hline
   Exact          &  $4.244$      & $1.183$      &  $1.793$     & $1.268$     &  $98.5$\% \\
   ph-QMC         &  $4.242(8)$   & $1.182(6)$   &  $1.799(1)$  & $1.262(3)$  &  $98.4$\% \\
   GP             &  $4.429$      & $1.029$      &  $1.800$     & $1.599$     & $100.0$\% \\
  \hline
  \hline
  \end{tabular}

\end{center}
\end{table}

We now show calculations on a large system with realistic $a_s$ values.
We use the unconstrained QMC (u-QMC) as the reference.
For weak to moderate interaction strength, the unconstrained QMC can be
carried out for a short period of time $\tau$ before the signal is
completely lost in large Monte Carlo fluctuations.
To obtain the desired accuracy, we perform many short QMC runs
and average the results.
For each scattering lengths, we verified that the short runs have reached
convergence with respect to the projection time.
The severity of the phase problem grows rapidly with $a_s$,
and such runs are not possible for large values of $a_s$.

Table~\ref{tbl:T3D+phls-benchmark-as80-500} shows
the phaseless QMC with the local-energy approximation
[\Eq{eq:iz-prop-weight-EL}]
for 3D trapped gas of $100$ atoms with $a_s = 80\,\Angstrom$
and $500\,\Angstrom$.
The first case represents a typical situation
in the trapped atomic gas experiments far from Feshbach resonances,
while the second is a medium-strength interaction deep into the range
of $a_s$ we study.
The $\Dt$ parameter was adjusted so that the Trotter error is
similar to or smaller than the statistical error.
\begin{table}[!htbp]
\begin{center}

  \caption{\label{tbl:T3D+phls-benchmark-as80-500}
Benchmark of QMC calculations with and without the phaseless constraint
for $a_s = 80\,\Angstrom$ and $500\,\Angstrom$.
We simulate 100 atoms in a 3D harmonic trap with
$\aho = 8546\,\Angstrom$.
The simulation lattice is $24 \times 24 \times 24$.
Shown here are per-particle quantities.
All energies are expressed in the unit of $\hbar\omega\_0$. 
}
  \begin{tabular}{lllll}
  \hline
  \hline
   Type                  &  $E / N$      & $\kinXV / N$ &  $\trapXV / N$ & $\potXV / N$ \\
  \hline
   $a_s = 80\,\Angstrom$ \\
   ph-QMC         &  $1.7943(3)$  & $0.5984(3)$  &  $0.96029(9)$  & $0.23562(8)$ \\
   u-QMC                 &  $1.7947(2)$  & $0.5987(2)$  &  $0.96006(4)$  & $0.23594(4)$ \\
   GP                    &  $1.7924$     & $0.5947$     &  $0.95649$     & $0.24121$    \\
\COMMENTED{
  \hline
   $a_s = 300\,\Angstrom$ \\
   ph-QMC ($\EL$)        &  $2.321(2)$   &  $0.503(2)$  &  $1.3190(4)$   &  $0.4995(6)$ \\
   ph-QMC                &  $2.314(2)$   &  $0.495(2)$  &  $1.3200(4)$   &  $0.4991(6)$ \\
   u-QMC                 &  $2.321(1)$   &  $0.500(1)$  &  $1.3185(4)$   &  $0.5032(5)$ \\
   GP                    &  $2.296$      &  $0.460$     &  $1.2857$      &  $0.5503$    \\
}
  \hline
   $a_s = 500\,\Angstrom$ \\
   ph-QMC        &  $2.6777(2)$  &  $0.500(3)$  &  $1.5638(6)$   &  $0.591(1)$  \\
   u-QMC                 &  $2.6811(4)$  &  $0.511(7)$  &  $1.563(2)$    &  $0.614(3)$  \\
   GP                    &  $2.620$      &  $0.408$     &  $1.4901$      &  $0.721$     \\
  \hline
  \hline
  \end{tabular}

\end{center}
\end{table}
We see that the agreement between the phaseless and unconstrained calculations is
good.

As a further check, we compare our QMC result on the uniform Bose gas with
an earlier diffusion Monte Carlo (DMC) calculation by Giorgini and
co-workers\Cite{Giorgini1999}, which is exact.
We use their results for the soft sphere potential with large radius of
$R = 5 a_s$, which best matches our situation, namely
$\varsigma \sim 2 R \sim 10 a_s$. As we show in
the left panel of \Fig{fig:Uscan-as.100p.D=6.66-E+condfrac},
our results agree well with their DMC energies.

\section{Bogoliubov ground state}
\label{app:Bog-gs}

The Bogoliubov approximation for the \emph{homogenous Bose gas}
assumes a macroscopic occupancy of the lowest
energy state ($\kvec = \mathbf{0}$), namely $(N - N_0) \ll N$.
We will work in the thermodynamic limit,
$N \rightarrow \infty$ and $\Omega \rightarrow \infty$, keeping the
density $\rho = N / \Omega$ finite.
The creation and annihilation operators for the zero-momentum state are
approximated as scalars,
\eql{eq:Bog-a0}
{
    \Cvarphi{}(\mathbf{0})
&   \approx
    \Dvarphi{}(\mathbf{0})
    \approx
    \sqrt{N_0}
    \,.
}
We then ignore all terms higher than quadratic in the remaining creation
and annihilation operators.
The form of the two-body potential also requires that $k \as \ll 1$.
Within this approximation, the energy per particle is given by\Cite{Huang1987}
\eqsl{
    \label{eq:Bog-E-series}
    \widetilde{E}_{\mathrm{Bog}}
&   \equiv
    E_{\mathrm{Bog}}/N
\\
& = \frac{4\pi\rho\as}{2N}
    \left[N
        - \sum_{\kvec \neq 0}
          \left(
                \alpha_{\kvec}^2 - \frac{1}{2x_{\kvec}^2}
          \right)
    \right]
\\
    \label{eq:Bog-E}
& = {2\pi \rho a_s}
    \left( 1 + \frac{128}{15\sqrt{\pi}} \sqrt{\rho a_s^3} \right)
    \,,
}
and the occupation of the $\kvec$ momentum state by~%
\footnote{
The continuum momentum distribution, which is often referred to in the
main text, \emph{per particle}, is given by
$
    {\pi}(\kvec)/N
    \equiv
    \widetilde{\pi}(\kvec)
  = n(\kvec) / [(2\pi)^3 \rho]
$.
}
\eql{eq:Bog-pdensity}
{
    n\_{\mathrm{Bog}}(\kvec)
  = \frac{\alpha_{\kvec}^2}{1 - \alpha_{\kvec}^2}
    \qquad
    \textrm{($\kvec \neq \mathbf{0}$)}
    \,,
}
where
\eqsl{
    \label{eq:Bog-x}
    x_{\kvec}
&   \equiv
    \frac{|\kvec|}{({8\pi\rho\as})^{1/2}}
    \equiv
    \xi |\kvec|
    \,,
\\
    \label{eq:Bog-alpha}
    \alpha\_{\kvec}
&   \equiv
    1 + x_{\kvec}^2 - x_{\kvec} \sqrt{x_{\kvec}^2 + 2}
    \,.
}
The quantity $\xi \equiv (8\pi\rho\as)^{-1/2}$ is the healing
length\Cite{Leggett01} of the condensate.
The condensate fraction is given by
\eqsl{
    \label{eq:Bog-cond-depl-sum}
    \frac{N_0}{N}
& = 1 - \frac{1}{N} \sum_{\kvec \neq \mathbf{0}} n(\kvec)
\\
    \label{eq:Bog-cond-depl2}
& = 1 - \frac{8}{3} \sqrt{\frac{\rho\as[3]}{\pi}}
    \,.
}

The kinetic energy per particle can be computed through
\eql{eq:Bog-KE}
{
    \widetilde{T}\_{\mathrm{Bog}}
  = \frac{1}{2N} \sum_{\kvec \neq 0}
    |\kvec|^2 n(\kvec)
    \,.
}
The summation, however, must be performed with care,
as mentioned in \Sec{ssec:Bogoliubov}.
The analytic results for the energy and condensate fraction,
\Eqs{eq:Bog-E} and \eqref{eq:Bog-cond-depl2}, are obtained by extending
the summation variable to infinity, because the contribution from outside
the $k \as \ll 1$ region is assumed to be small.
This assumption does not hold for the kinetic energy, since the sum
diverges due to the unphysical nature of $n(\kvec)$ at large $|\kvec|$.


To benchmark our Bogoliubov approach, we perform QMC and Bogoliubov
calculations for a homogenous Bose gas at different scattering lengths, as
shown in \Fig{fig:Uscan-as.100p.D=6.66-E+condfrac}.
We compute the energetics and condensate fraction using three different
methods: GP, Bogoliubov, and QMC.
As we see here, the first-order Bogoliubov approximation estimates the
energetics and condensate fraction very well for a small enough gas
parameter (here $\rho\as[3] \lesssim 10^{-4}$).

\begin{figure}[hbtp]

  \includegraphics[scale=1.4]{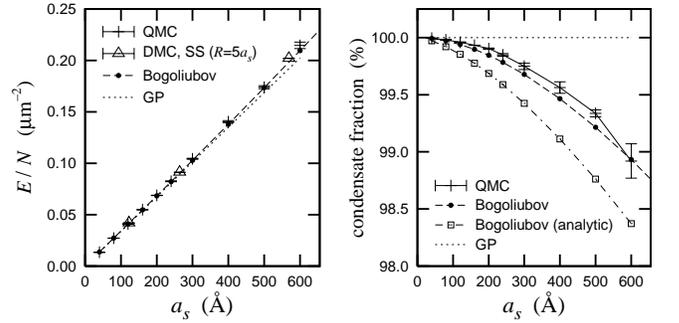}
  \caption{\label{fig:Uscan-as.100p.D=6.66-E+condfrac}%
Benchmark of our QMC and Bogoliubov-LDA calculations in
\emph{the uniform Bose gas}.
The system is the same as that in \Fig{fig:Uscan-as.100p.D=6.66-K+V2b}.
The upper triangle data points in the g.s.{} energy plot are from diffusion
Monte Carlo (DMC) calculations using a soft sphere (SS)
potential\Cite{Giorgini1999}.
In the condensate fraction plot, we also show the analytic Bogoliubov
result without truncation of $|{\mathbf k}|$, \Eq{eq:Bog-cond-depl2}.
}

\end{figure}

Note that the condensate fraction estimated by Bogoliubov with the
truncation in the sum in $k$-space agrees much better with QMC than
the analytic Bogoliubov.
The analytic Bogoliubov estimate is off, as discussed above,
because it is extrapolated
to an infinite box size, and it includes contributions from very high
momentum states.

We note that the kinetic energy,
which is very small in the small $a_s$ regime, is no
longer negligible for larger $a_s$ values.
For $a_s = 600\,\Angstrom$, or equivalently
$\rho a_s^3 = 1.2 \times 10^{-4}$,
the kinetic energy (see \Fig{fig:Uscan-as.100p.D=6.66-K+V2b})
is about $37\%$ of the total energy. This is consistent with
our discussion in \Sec{sec:Results} on the balance between
the mean-field and correlation effects.
%

We can extend the Bogoliubov analysis above to deal with the case of a
\emph{inhomogeneous, trapped gas}.
We use the so-called local-density approximation (LDA) by treating each
lattice site as a locally homogenous gas.
The density profile $\rho(\rvec)$ can be estimated using GP or any other
methods which provides a good approximation to the density profile.
Using the same $k$-space lattice as QMC, we compute the ``local''
energetics (per particle) and condensate fraction.
The density is then used to weight-average the local contributions.
The Bogoliubov-LDA estimate of the kinetic energy is
\eql{eq:LDABog-KE2}
{
    \XV{\Top}\_{\mathrm{Bog}\textrm{-}\mathrm{LDA}}
&=  -\Half \int d^3\rvec \,
    \sqrt{\rho(\rvec)} \, \nabla^2 \sqrt{\rho(\rvec)}
\\
&+~ 
    \int d^3\rvec \,
    \tilde{T}_{\mathrm{Bog}}[\rho(\rvec)]\rho(\rvec)
    \,.
}
The interaction energy is given by
\eql{eq:LDABog-V2B}
{
    \potXV\_{\mathrm{Bog}\textrm{-}\mathrm{LDA}}
& = \int d^3\rvec
    \left( \widetilde{E}\_{\mathrm{Bog}}[\rho(\rvec)]
         - \widetilde{T}[\rho(\rvec)]\_{\mathrm{Bog}}
    \right)
    \rho(\rvec)
    \,.
}
The external trap energy is straightforward to compute, namely
\eql{eq:LDABog-Vtrap}
{
    \trapXV\_{\mathrm{Bog}\textrm{-}\mathrm{LDA}}
  = \int d^3\rvec \, V_{\mathrm{trap}}(\rvec) \rho(\rvec)
    \,.
}


\bibliography{AFQMC-bib-entries}    

\end{document}